\documentclass[fleqn,twoside]{article}
\usepackage{espcrc2}

\usepackage{graphicx}
\usepackage[figuresright]{rotating}

\newcommand{\AmS}{{\protect\the\textfont2
  A\kern-.1667em\lower.5ex\hbox{M}\kern-.125emS}}
\def\be{\begin{equation}}
\def\ee{\end{equation}}
\def\bea{\begin{eqnarray}}
\def\eea{\end{eqnarray}}
\def\ppg{\pi^{+}\pi^{-}\gamma}
\def\mmg{\mu^{+}\mu^{-}\gamma}
\def\eeg{e^{+}e^{-}\gamma}

\title{Measurement of hadronic cross section and preliminary results on the pion form factor using the radiative return
at DA$\Phi$NE} 

\author{The KLOE Collaboration~\thanks{
The KLOE Collaboration: A.~Aloisio,
F.~Ambrosino,
A.~Antonelli,
M.~Antonelli,
C.~Bacci,
G.~Bencivenni,
S.~Bertolucci,
C.~Bini,
C.~Bloise,
V.~Bocci,
F.~Bossi,
P.~Branchini,
S.~A.~Bulychjov,
R.~Caloi,
P.~Campana,
G.~Capon,
G.~Carboni,
M.~Casarsa,
V.~Casavola,
G.~Cataldi,
F.~Ceradini,
F.~Cervelli,
F.~Cevenini,
G.~Chiefari,
P.~Ciambrone,
S.~Conetti,
E.~De~Lucia,
G.~De~Robertis,
P.~De~Simone,
G.~De~Zorzi,
S.~Dell'Agnello,
A.~Denig,
A.~Di~Domenico,
C.~Di~Donato,
S.~Di~Falco,
A.~Doria,
M.~Dreucci,
O.~Erriquez,
A.~Farilla,
G.~Felici,
A.~Ferrari,
M.~L.~Ferrer,
G.~Finocchiaro,
C.~Forti,
A.~Franceschi,
P.~Franzini,
C.~Gatti,
P.~Gauzzi,
S.~Giovannella,
E.~Gorini,
F.~Grancagnolo,
E.~Graziani,
S.~W.~Han,
M.~Incagli,
L.~Ingrosso,
W.~Kluge,
C.~Kuo,
V.~Kulikov,
F.~Lacava,
G.~Lanfranchi,
J.~Lee-Franzini,
D.~Leone,
F.~Lu,
M.~Martemianov,
M.~Matsyuk,
W.~Mei,
L.~Merola,
R.~Messi,
S.~Miscetti,
M.~Moulson,
S.~M\"uller,
F.~Murtas,
M.~Napolitano,
A.~Nedosekin,
F.~Nguyen,
M.~Palutan,
L.~Paoluzi,
E.~Pasqualucci,
L.~Passalacqua,
A.~Passeri,
V.~Patera,
E.~Petrolo,
L.~Pontecorvo,
M.~Primavera,
F.~Ruggieri,
P.~Santangelo,
E.~Santovetti,
G.~Saracino,
R.~D.~Schamberger,
B.~Sciascia,
A.~Sciubba,
F.~Scuri,
I.~Sfiligoi,
T.~Spadaro,
E.~Spiriti,
G.~L.~Tong,
L.~Tortora,
E.~Valente,
P.~Valente,
B.~Valeriani,
G.~Venanzoni,
S.~Veneziano,
A.~Ventura,
G.~Xu,
G.~W.~Yu.
} 
\\presented by G. Venanzoni 
\address{Universit\`a and Sezione INFN Pisa}}

\begin{document}

\begin{abstract}
In the fixed energy environment of the 
$e^{+}e^{-}$ collider DA$\Phi$NE, 
KLOE can measure the cross section of the process 
$e^{+}e^{-} \rightarrow$ hadrons as a function of the hadronic 
system energy using the radiative return. At energies below 1 GeV, 
$e^{+}e^{-} \rightarrow \rho \rightarrow \pi^{+}\pi^{-}$ 
is the dominating hadronic process.  We report here on
the status of the analysis for the  $e^{+}e^{-} \rightarrow \ppg$ channel,  
which allows to obtain a preliminary measurement 
of the pion form factor using  an integrated luminosity of $\sim73 \,pb^{-1}$. 
\vspace{1pc}
\end{abstract}

\maketitle

\section{Measuring the hadronic cross section with the KLOE detector}

\subsection{Motivation}

The measurement of the hadronic cross section at low energy
is of great importance for the improvement of the theoretical 
error of the anomalous magnetic moment of the muon, 
$a_\mu = (g_\mu-2)/2$.
The hadronic contribution $a_\mu^{hadr}$
is given by the hadronic vacuum polarization and 
cannot be calculated at low energy in the framework of 
perturbative QCD. Following a phenomenological approach, 
the hadronic contribution can however be evaluated 
from the measurement of R through a dispersion relation.

\begin{equation}
a_{\mu}^{hadr} = (\frac{\alpha m_{\mu}}{3 \pi})^2 \int_{4
m_{\pi}^2}^{\infty} ds \frac{R(s) \hat{K}(s)}{s^2}   ,\\
\label{amu}
\end{equation}
where $R(s)=\frac{\sigma(e^+ e^- \rightarrow
hadrons)}{\frac{4\pi\alpha^2(s)}{3s}}$ and  the kernel 
$\hat{K}(s)$ is a smooth bounded function growing from
$0.63$ at threshold to $1$ at $\infty$. Due to the $1/s^2$ dependence in
the integral, hadronic data at low energies are
strongly enhanced in the contribution to $a_\mu^{hadr}$.
The error of the hadronic contribution is therefore 
given by the limited knowledge of hadronic cross section data.
This error is the dominating contribution to the total error of
$a_\mu^{theo}$ ($\delta a_\mu^{theo} \approx \delta a_\mu^{hadr}$)~\cite{Czarnecki:2001pv}
\\

The discrepancy between the theoretical 
calculation~\cite{Czarnecki:2001pv,DeTroconiz:2001wt,deRafael:2002xy,Hagiwara:2002ma}, 
and the recent experimental value of $a_\mu$~\cite{Bennett:2002jb} 
depends if $\tau$ data are used  or not in the evaluation of 
$a_\mu^{had}$~\cite{Jegerlehner:2001wq,Davier:1998si}; 
a recent re-evaluation of the hadronic contribution to $a_\mu$~\cite{Davier:2002dy}  found:
$a_\mu^{theo} -   a_\mu^{exp}$ = $(33.9 \pm 11.2)\times 10^{-10}$ $[e^+e^- \,based]$,
$a_\mu^{theo} - a_\mu^{exp} $  =
 $(16.7 \pm 10.7)\times 10^{-10}$ $[\tau\, based]$, corresponding to 3.0 
and 1.6 standard deviations respectively,  reflecting also  
a $1.6\,\sigma$  discrepancy between  $e^+e^-$ and $\tau$ evaluations of $a_\mu^{had}$.
A discussion on this subject can also be found in~\cite{davier}.

This unclear situation makes a precise measurement of  
the hadronic cross 
section at low energy mandatory, particularly in the channel 
 $e^{+}e^{-} \rightarrow \rho \rightarrow \pi^{+}\pi^{-}$, which is the main
 ingredient for $a_\mu^{had}$. 
Such a cross section has been already measured by the CMD-2 collaboration with an energy scan in the range below 
1.4 GeV~\cite{eidelman}; an accuracy of 0.6\% was recently achieved
in the region between 0.61 and 0.96 GeV~\cite{Akhmetshin:2001ig}.

\subsection{Radiative Return}
DA$\Phi$NE~\cite{dafne} is an $e^{+}e^{-}$ storage-ring collider 
working at the $\phi$ resonance (1020 MeV). As an experiment 
at a collider with a fixed centre of mass energy, KLOE can 
measure the hadronic cross section $\sigma$($e^{+}e^{-} \rightarrow$ hadrons)
 as a function of the hadronic system energy using the 
{\em radiative return}~\cite{Spagnolo:1998mt,Binner:1999bt}
i. e. studying the process $e^{+} e^{-} \rightarrow$ hadrons$+\gamma$.
 The emission of one photon before the beams interact 
(Initial State Radiation, ISR in the following) lowers the interaction energy and 
 makes possible to produce the hadronic system with an invariant 
mass varying from the $\phi$ mass down to the production threshold. 

The method represents an alternative approach to the conventional
 energy scan used so far for hadronic cross section measurements.
 A very solid theoretical understanding of Initial State Radiation 
(described by the radiation function $H$) is mandatory in order to 
extract the cross section $\sigma$($e^{+}e^{-} \rightarrow$ hadrons) 
as a function of $Q_{had}^{2}$ from the measured differential cross section
 $d\sigma$($e^{+}e^{-} \rightarrow$ hadrons$+\gamma$)$/dQ_{had}^{2}$:
\begin{eqnarray*}
Q_{had}^{2} \cdot \frac{d\sigma(e^{+}e^{-} \rightarrow {\rm hadrons} + \gamma)}{dQ_{had}^{2}}& =& \\
 \sigma(e^{+}e^{-} \rightarrow {\rm hadrons}) \cdot H(Q_{had}^{2},\theta_{\gamma}) & & \\
\label{eq:radiation}
\end{eqnarray*}
where $Q_{had}^{2}$ and $\theta_{\gamma}$
 are respectively the invariant mass squared of the hadronic system and the acceptance cut
 on the polar angle of the photon. 

Radiative corrections for $e^{+}e^{-} \rightarrow \pi^{+}\pi^{-}\gamma$, 
 have been calculated up to ISR NLO by different
 theoretical groups~\cite{Arbuzov:1998te,Konchatnij:1999xe,Khoze:2000fs,Khoze:2002ix,Rodrigo:2001kf,Kuhn:2002xg}, and 
recently implemented in a new Monte Carlo generator named PHOKHARA~\cite{Rodrigo:2002hk,rodrigo},
 This generator 
with a claimed accuracy of 0.5\% has been used in the present analysis.

  We want to stress that 
the radiative return has the merit compared 
with the conventional energy scan that the systematics
 of the measurement (e.g. normalization, beam energy)
 are the same for any experimental point and must not 
be evaluated at each energy step. However a precise determination of the angle
of the hard photon as well as the full control of events with the photon 
emitted in the final state (pure QED FSR or other resonant processes,
 as for example $\phi\to\pi^+\pi^-\gamma$~\cite{Melnikov:2000gs}), is required.

\section{Analysis of the process $e^{+}e^{-} \rightarrow \ppg$}
Due to the importance of the $2\pi$ final state for $a_\mu^{had}$, 
we concentrated at 
KLOE on the analysis of the radiative process $e^{+}e^{-} \rightarrow \ppg$, 
whose feasibility has been extensively studied by Monte Carlo~\cite{Cataldi:1999dc}.

\subsection{The KLOE Detector}
KLOE \cite{Aloisio:1993fy} is a typical $e^+ e^-$ multiple purpose detector 
with cylindrical geometry,
consisting of a large helium based drift chamber (DC,
\cite{dc}),
surrounded by an electromagnetic calorimeter
(EmC, \cite{emc}) and a superconducting magnet ($B=0.52$ T). 
The detector has been
designed for the measurement of $CP$ violation in the neutral kaon system,
i.e. for precise detection of the decay products of $K_S$ and $K_L$.
These are low momenta charged tracks 
($\pi^\pm, \mu^\pm, e^\pm$ with a momentum range from $150 $ MeV/c 
to $270 $ MeV/c) and low energy photons (down to $20$ MeV). 
\\
The DC dimensions ($3.3$ m length, $2$ m radius), 
the drift cell shapes ($2$x$2$ cm$^2$ cells for the inner $12$ 
layers, $3$x$3$ cm$^2$ cells for the outer $46$ layers) and the choice of 
the gas mixture ($90\%$ Helium, $10\%$ Isobutane; $X_0 = 900$ m) 
had to be optimized for the requirements prevailing at  
a $\phi$ factory. The KLOE design results in a very good momentum 
resolution: $\sigma_{p_{\bot}} / p_{\bot} \leq 0.4\%$ at high
tracking efficiencies ($> 99\%$).
\\
The EmC is made of a matrix of scintillating fibres embedded 
in lead, which guarantees a good energy re\-solution 
$\sigma_E / E = 5.7 \% / \sqrt{E({\rm GeV})}$ and excellent 
timing resolution $\sigma_t = 57 {\rm ps} / \sqrt{E({\rm GeV})} \oplus 50$ ps.
The EmC consists of a barrel and two endcaps  
which are surrounding the cylindrical DC; this gives a 
hermetic coverage of the solid angle ($98\%$). 
However, the acceptance of the EmC below $\approx 20^\circ$ is 
reduced due to the presence of quadrupole magnets 
close to the interaction point and does not allow to 
measure e.g. the photon of $\pi^+ \pi^- \gamma$ events with low 
$\theta_\gamma$ angles. 
\begin{figure}
\includegraphics[width=17pc]{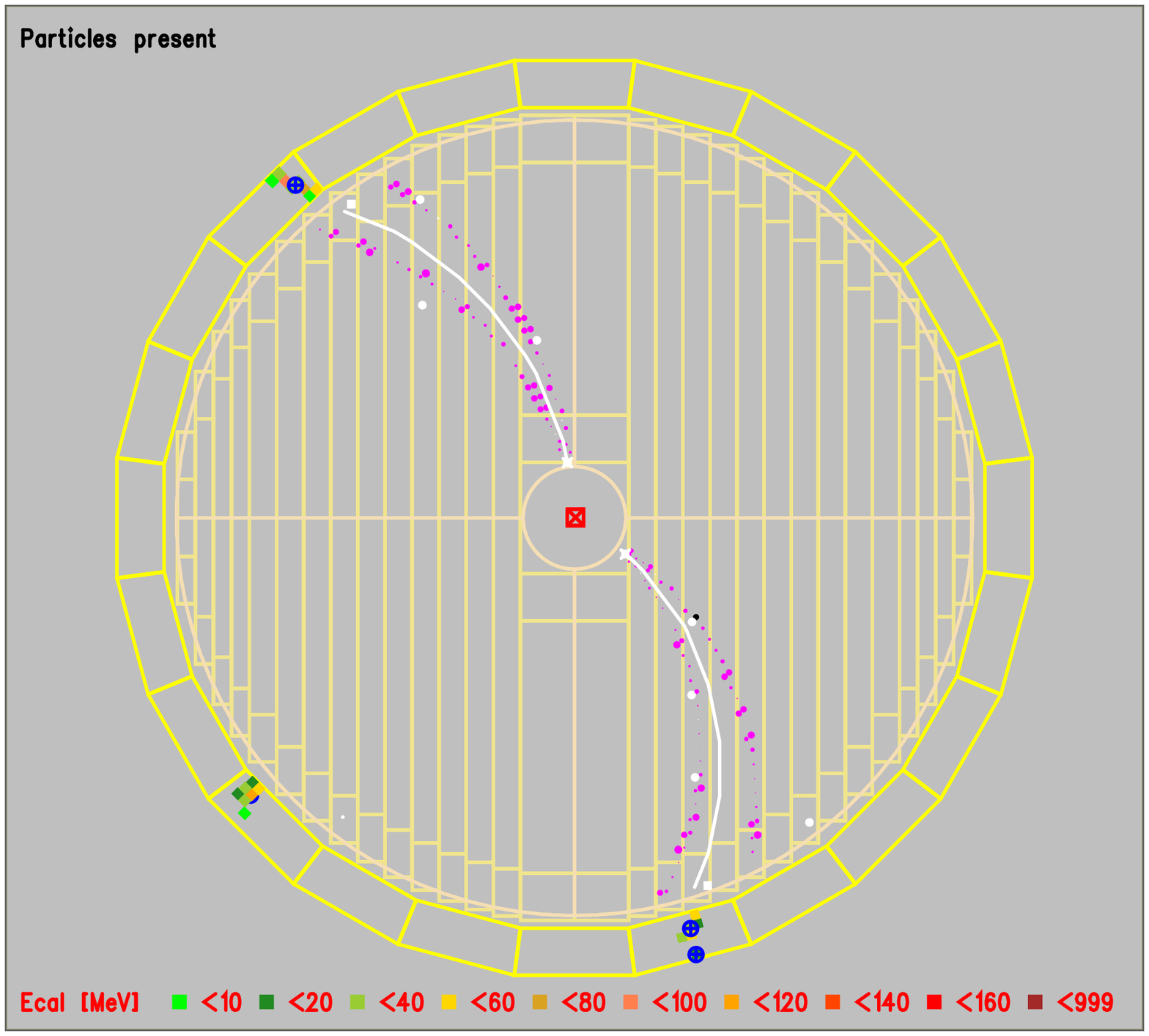}
\caption{KLOE event display of a $\ppg$ event.}
\label{fig:ppg}
\end{figure}

It will be shown in the following that an efficient
selection of the $\pi^+ \pi^- \gamma$
signal is possible, without requiring 
an explicit photon detection. 
The relatively simple signature of the signal
(2 high momentum tracks from the interaction point, see 
Fig.~\ref{fig:ppg}) and  
the good momentum resolution of the KLOE tracking detector
allow us to perform such a selection.

\subsection{Signal selection}\label{subsec:sig_sel}
The selection of $e^{+}e^{-} \rightarrow \ppg$ events is done in the following steps:
\begin{itemize} 
\item {\it detection of two charged tracks}, with polar angle bewteen $40^o$
and $140^o$, coming from a vertex in the fiducial volume $R<8$ cm, $|z|< 15$ cm. The cuts on the transverse momentum $p_{T} > 200$ MeV or on the longitudinal momentum $|p_{z}| > 90$ MeV reject tracks spiralizing along the beam line, ensuring good reconstruction conditions. 
The probability to reconstruct a vertex in the drift chamber is $\sim 95\%$ and has been studied with Bhabha data, 
selected using the calorimeter only. The overall tracking reconstruction efficiency has been also 
evaluated using $\pi^{+}\pi^{-}\pi^{0}$ events selected by detecting $\pi^{0}$ in the electromagnetic calorimeter; 
\begin{figure}
\includegraphics[width=17pc]{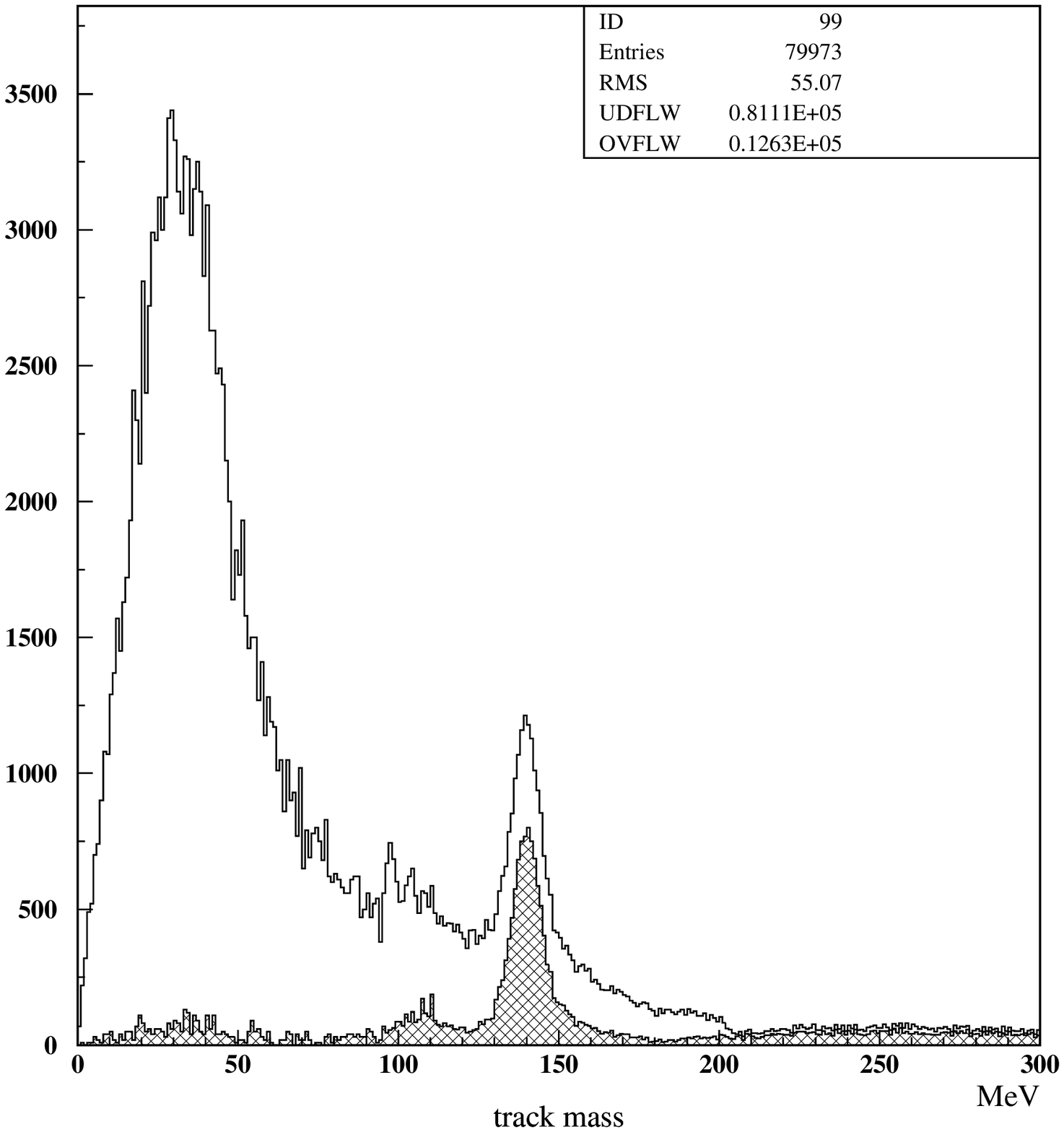}
\includegraphics[width=18pc]{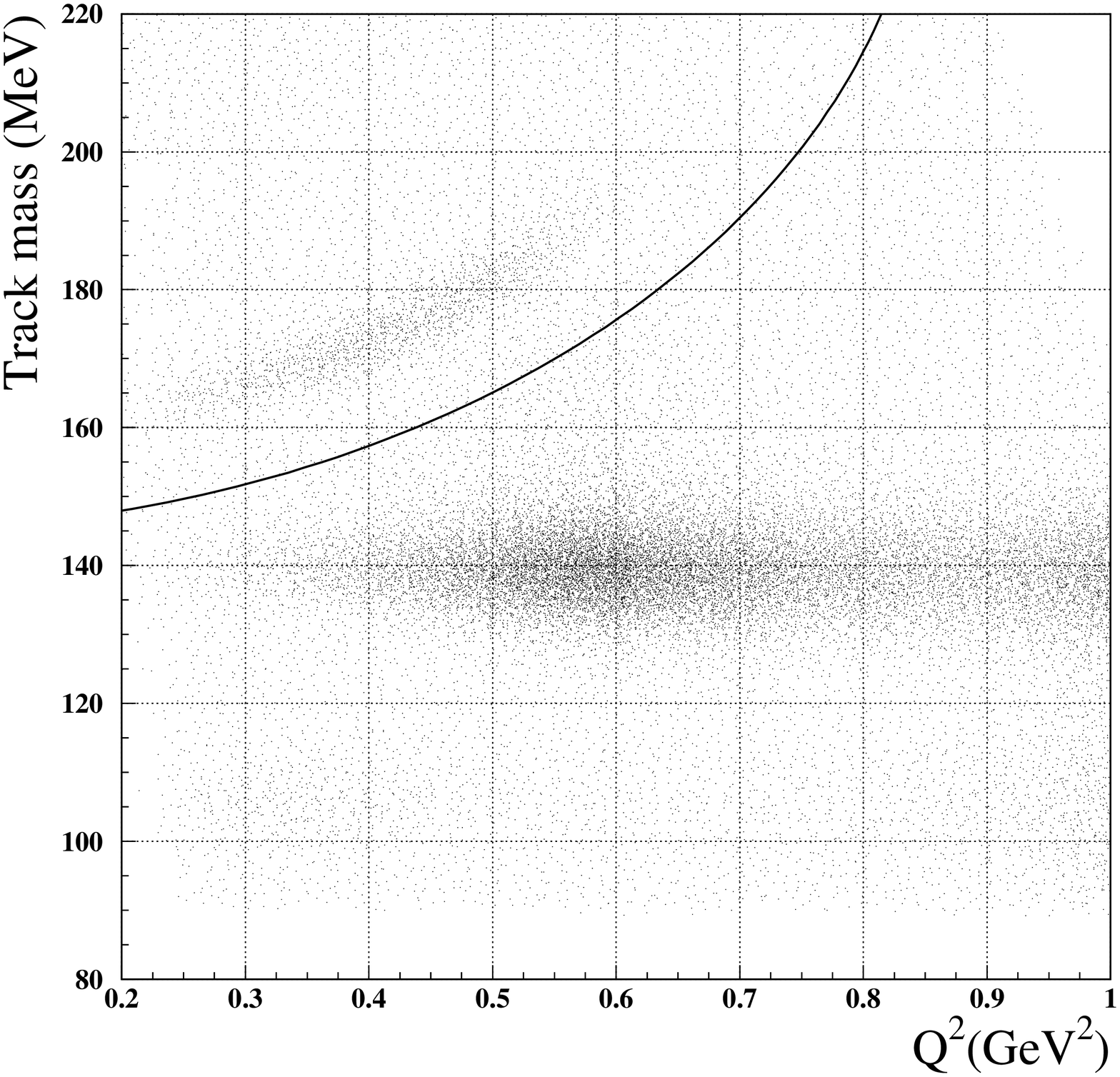}
\caption{{\it Up}: track mass distribution before and after the likelihood selection. 
$\mmg$, $\ppg$ distributions are peaked at the proper mass, three pion events populate 
the region on the right of the $\ppg$ peak. {\it Down}: cut in the plane 
($Q_{\pi\pi}^{2}$, track mass) used to reject $\pi^{+}\pi^{-}\pi^{0}$ events.} 
\label{fig:trkmass}
\end{figure}

\item {\it identification of pion tracks}: a Likelihood Method, 
using the time of flight of the particle and the shape of the energy deposit in the electromagnetic calorimeter, 
has been developed to reject the $e^{+}e^{-} \rightarrow \eeg$ background.
 A control sample of $\pi^{+}\pi^{-}\pi^{0}$ has been used to study the behaviour of pions in the electromagnetic 
calorimeter, since their  interaction are not well reproduced by Monte Carlo.  
The effect of the selection on $\ppg$ and $\eeg$ events is visible in the 
upper plot of Fig.~\ref{fig:trkmass}, 
as a function of the track mass, $M_{track}$: this variable is calculated from the 
reconstructed momenta, $\vec{p}_{+}$, $\vec{p}_{-}$, applying 4-momentum 
conservation, under the hypothesis that the final state consists of two particles with the same mass and one photon. 
For each event class ($\mmg$, $\ppg$) the track mass distribution is peaked at the proper mass; 
$\pi^{+}\pi^{-}\pi^{0}$ events populate the region on the right of the $\ppg$ peak. 
As it can be noted from  Fig.~\ref{fig:trkmass}, {\it up}, $\eeg$ events are drastically reduced, while the 
$\ppg$ peak is essentially unaffected by this selection 
(the estimated signal efficiency is larger than 98\%).
\item {\it cut on the track mass}: $\mmg$ events are rejected by a cut at 120 MeV in the track mass. The discrimination between $\pi$ and $\mu$ using calorimeter information is not helpful since pions behave frequently like minimum ionizing particles. After this cut we find a contamination of $\mmg$ background smaller than $1\%$. 
$\pi^{+}\pi^{-}\pi^{0}$ events are rejected with a cut in the two-dimensional distribution of the track mass 
versus the two pion invariant mass squared, $Q_{\pi\pi}^{2}$, shown in Fig.~\ref{fig:trkmass}, {\it down}.
 Due to the large production cross section for these events ($BR(\phi \rightarrow \pi^{+}\pi^{-}\pi^{0})\sim 15\%$,
 $\sigma\sim500$ nb) and the similar kinematics, some residual contamination is 
 expected at small $Q_{\pi\pi}^{2}$ values ($Q_{\pi\pi}^{2} <0.4\, GeV^2$).  
The efficiency of the track mass cut, as evaluated from Monte Carlo, is $\sim 90\%$;
\item {\it definition of the angular acceptance:} 
The polar angle of the photon, $\theta_{\gamma}$ is calculated 
from the charged tracks
as $180^0-\theta_{\pi\pi}$, where $\theta_{\pi\pi}$ is the polar angle of the two pion system. 
Two fiducial volumes for the analysis of 
$\ppg$ events are defined: $\theta_{\gamma} < 15^{o}$ 
or $\theta_{\gamma} > 165^{o}$ (small angle analysis) and 
$55^{o}< \theta_{\gamma} < 125^{o}$ (large angle analysis). These two regions
 mainly differ for FSR and background contamination~\cite{Valeriani:2002yk}.
\end{itemize}
In this paper we will show the preliminary 
results obtained on the analysis at small angle ($\theta_{\gamma} < 15^{o}$ 
or $\theta_{\gamma} > 165^{o}$, $40^{o}< \theta_{\pi} < 140^{o}$). 

\section{Final State Radiation}
It is now a common understanding that the radiation of one hard photon by final pions 
 should be included in the measurement of  
the hadronic cross section  for the evaluation of 
$a_{\mu}^{had}$~\cite{Melnikov:2001uw,DeTroconiz:2001wt,Hoefer:2001mx,eidelman,Akhmetshin:2001ig}.

However~\cite{Binner:1999bt}, in the case of radiative return,   
the differential cross section for the process $e^{+}e^{-} \rightarrow \ppg$
is proportional to the pion form factor squared at the measured $Q^2_{\pi\pi}$ only for ISR:  
\be
\left (\frac {{\rm d}{\sigma}}{{\rm d} Q^2} \right )_{\rm ISR}
\sim |F_\pi(Q^2)|^2 \sim \sigma_{e^+e^- \to \pi^+ \pi^-}(Q^2),
\ee
while for FSR the differential cross section~\footnote{If no additional photons are emitted in the initial state.} 
is proportional  to the pion form factor evaluated at the $\phi$ mass squared:
\be
\left ( \frac {{\rm d}{\sigma}}{{\rm d} Q^2} \right )_{\rm FSR} 
\sim |F_\pi(M_{\phi}^2)|^2 \ne \sigma_{e^+e^- \to \pi^+ \pi^-}(Q^2).
\ee

For this reason events due to FSR must be properly rejected by the 
analysis cuts.

\begin{figure} 
\includegraphics[width=17pc,height=14pc]{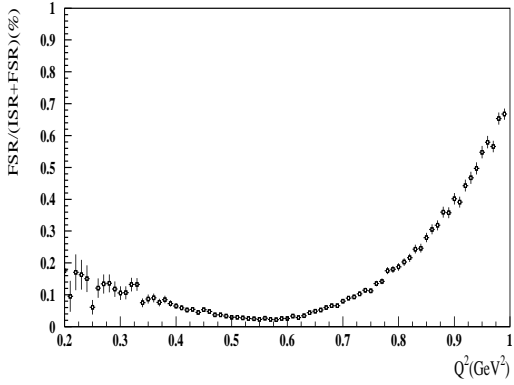}
\caption{Ratio $FSR/(FSR+ISR)$ as a function of the two pions invariant mass for 
$\theta_{\gamma}<15^o$ or $\theta_{\gamma}>165^o$, $40^o<\theta_{\pi}<140^o$.}
\label{fig:isrfsr}
\end{figure}

Fig.~\ref{fig:isrfsr} shows the ratio  $FSR/(ISR+FSR)$ as a function of the $Q^2_{\pi\pi}$ for the small angle region
($\theta_{\gamma}<15^o$ or $\theta_{\gamma}>165^o$, $40^o<\theta_{\pi}<140^o$). 
As it can be noted the average value of this ratio is below 0.5\%.

\section{Results}
\subsection{Effective cross section}
\begin{figure} 
\includegraphics[width=17pc]{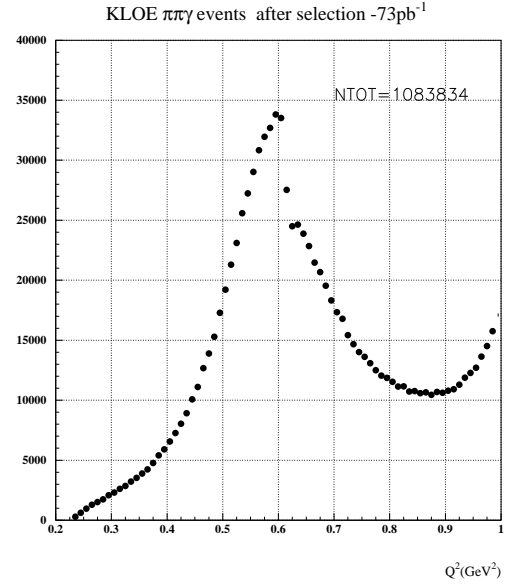}
\includegraphics[width=17pc]{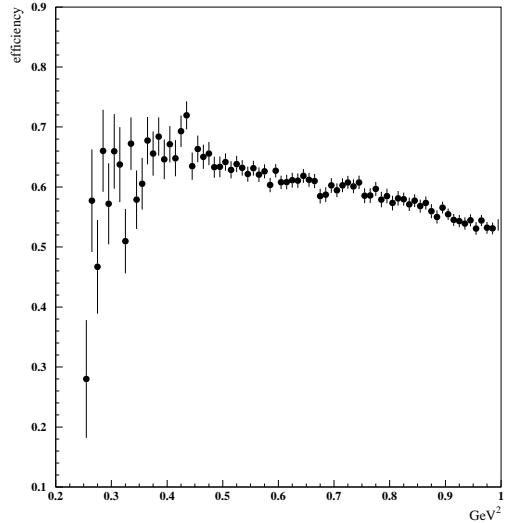}
\caption{{\it Up}: Number of $\ppg$ events selected by the analysis in the region 
$\theta_{\gamma}<15^o$ or $\theta_{\gamma}>165^o$, $40^o<\theta_{\pi}<140^o$.
{\it Down}: Global efficiency as a function of $Q^2_{\pi\pi}$.}
\label{fig:kloe}
\end{figure}

Fig.~\ref{fig:kloe} {\it up} shows the number of $\pi\pi\gamma$ events selected inside the small angle acceptance cuts
 at the end of the selection chain, for $100$ bins between $0.02$ and $1.02\,GeV^2$.
This distribution corresponds to $\sim \,73 pb^{-1}$ of analyzed data out of $\sim 200 \,pb^{-1}$ collected in 2001 
and already reconstructed. More than 1083000 were selected, corresponding to  $~15000$ events/$pb^{-1}$.

Even before unfolding the spectrum for the detector resolution effects, the $\rho-\omega$ interference 
(as well as the radiative tail) can be clearly seen, showing the excellent momentum resolution of the KLOE DC.
\subsection{Pion form factor calculation}
Neglecting FSR interference, the pion form factor can be extracted as a function of $Q^2_{\pi\pi}$,
from the observed $\ppg$ spectrum ($N^{obs}$) as:
\be
|F_\pi(Q^2_i)|^2 = \frac{N^{obs}_i}{\epsilon(Q^2_i) L H(Q^2_i)}
\label{eq:pion}
\ee
where  $ \epsilon(Q^2_i)$ is the global analysis efficiency, $L$ is the integrated luminosity and 
$H(Q^2_i)$ is the NLO cross section for $e^{+}e^{-} \rightarrow \ppg$ (only ISR) under the assumption of pointlike pions.

These quantities were evaluated in the following ways:
\begin{enumerate}
\item {\bf global efficiency}: 
 this is the product of trigger, reconstruction, filtering, likelihood 
and $M_{track}$ cut efficiency. Apart from the latter which was evaluated only by Monte Carlo, 
the other efficiencies were evaluated from unbiased 
samples of data with similar kinematics (like $\pi^+\pi^-\pi^0$, Bhabha's or $\ppg$ itself).
The Monte  Carlo  was used to generate the appropriate $\ppg$ phase space.

Fig.~\ref{fig:kloe}, {\it down}, shows the dependence of the global efficiency  
as a function of reconstructed $Q^2_{\pi\pi}$ (only statistical errors are shown).
Above $0.4\, GeV^2$ the average error is 2\%,  mainly dominated by the limited 
Monte Carlo statistics.
Below $0.4\, GeV^2$  the error is larger, reflecting again the  limited 
Monte Carlo statistics used in the evaluation of  $M_{track}$ cut efficiency.

\item{\bf Luminosity}:
The DA$\Phi$NE accelerator does not have luminosity monitors at small angle due to the existence of focusing quadrupole magnets very close to the interaction point. 
The luminosity is therefore measured using Large Angle Bhabhas ($55^{o} < \theta_{+,-} < 125^{o}$, $\sigma=425$nb). The number of LAB candidates are counted and normalized to the effective Bhabha cross section obtained from Monte Carlo. 

The precision of this measurement depends both on the understanding of experimental efficiencies and acceptances and on the theoretical knowledge of the process. 

The systematic errors arising from the LAB selection cuts are well below $1\%$. All the selection efficiencies concerning the LAB measurement (Trigger, EmC clusters, DC tracking) are above $98\%$ and well reproduced by the detector simulation. The background due to $\mmg$, $\ppg$ and $\pi^{+}\pi^{-}\pi^{0}$ is below $1\%$. 

KLOE uses two independent Bhabha event generators (the Berends/Kleiss generator~\cite{Berends:fs} 
modified for DA$\Phi$NE~\cite{Drago:1997px} and BABAYAGA~\cite{CarloniCalame:2000pz}). 

The very good agreement of the experimental distributions ($\theta_{+,-}$, $E_{+,-}$) 
with the event generators  and a cross check with an independent luminosity counter 
based on $e^{+}e^{-} \rightarrow \gamma \gamma(\gamma)$ indicates a precision of better than $1\%$. 

More systematics checks (e.g. the effect of a varying beam energy and of a displaced beam interaction point) are under way.

\item{\bf determination of H:}
The NLO cross section for pointlike pions $H$ was computed for each bin of $Q^2_{\pi\pi}$ using  a modified 
version of PHOKHARA, in which $F_\pi(Q^2)=1$.  

$2\cdot10^6$ events were generated within the 
acceptance of the small angle analysis:
$\theta_{\gamma}<15^o$ or $\theta_{\gamma}>165^o$, $40^o<\theta_{\pi}<140^o$. 
\begin{figure}[ht]
\includegraphics[width=17pc]{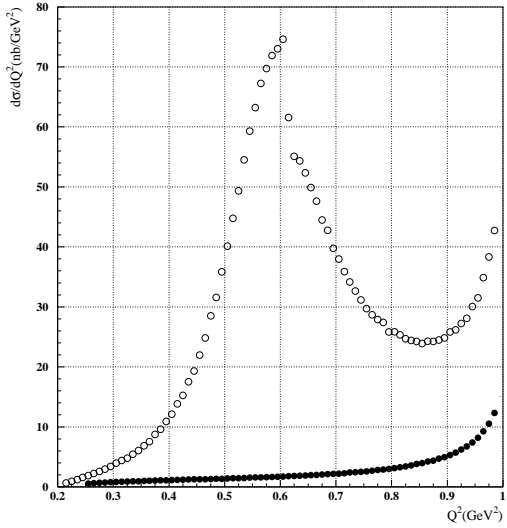}
\caption{Differential $\ppg$ NLO cross section from Monte Carlo, for  a given $F_\pi(Q^2)$ parametrization 
({\it white dots}), and assuming pointlike pions, {\it i.e.}  $F_\pi(Q^2)=1$ ({\it black dots}). The cuts are 
$\theta_{\gamma}<15^o$ or $\theta_{\gamma}>165^o$, $40^o<\theta_{\pi}<140^o$.}
\label{fig:htau02}
\end{figure}
Fig.~\ref{fig:htau02} shows a comparison between the differential cross section generated by Monte Carlo 
with a given parametrization of the pion form factor ({\it white dots}), and the same assuming pointlike pions
 ({\it black dots}).

The systematical error on the numerical evaluation of $H$ for the given acceptance cuts
 is estimated to be less than $1\%$ 
by comparing the pion form factor obtained by dividing  {\it bin-by-bin}  the two distributions of Fig.~\ref{fig:htau02} 
 with the expected  analytical parametrization. 
\end{enumerate}

\begin{figure}
\includegraphics[width=17pc]{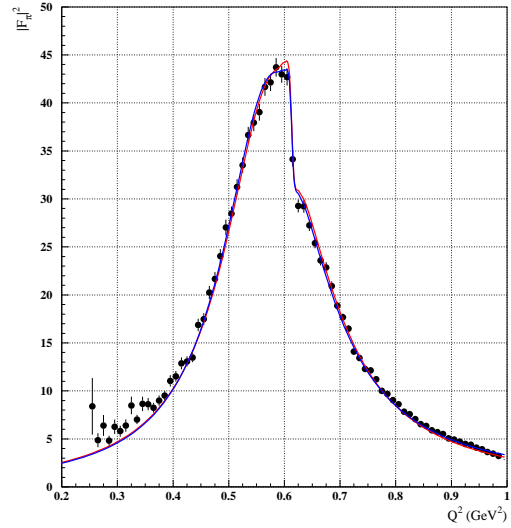}
\includegraphics[width=17pc]{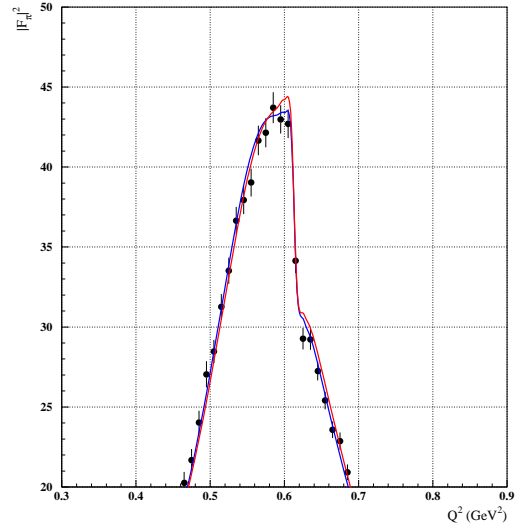}
\caption{{\it Up}: Preliminary measurement of the pion form factor. Overimposed are the analytical  $|F_\pi(Q^2)|^2$ 
parametrizations, obtained by a fit to the KLOE data using KS model ({\it blue line}),
 and the parametrization found from CMD-2 with the
GS model  ({\it red line}). The two curves are very similar. {\it Down}: zoom of $|F_\pi(Q^2)|^2$ around the $\rho$ peak.
The {\it blue line} is obtained by a fit to the KLOE data using KS model, while the {\it red} one is the
 parametrization found from CMD-2.}
\label{fig:pion}
\end{figure}
The pion form factor extracted using eq.~\ref{eq:pion},
 is shown in Fig.~\ref{fig:pion}. The spectrum 
can be divided in three regions according to the measurement 
 error: below $0.4\, GeV^2$ with an error of  5-10\%, between
$0.4\, GeV^2$ and $0.5\, GeV^2$, with an average error of 3\%, and above $0.5\, GeV^2$ with an 
average error of 2\%.  The measurement error is dominated
 by the limited Monte Carlo statistics used in the
evaluation of the efficiencies.  

\subsection{Fit of the pion form factor}
A preliminary fit was applied to data using 
the parametrization for $F_\pi(Q^2)$ found in ~\cite{Kuhn:1990ad} (KS):
\be
F_\pi(Q^2) = \frac{BW_{\rho}\frac{(1+\alpha BW_{\omega})}{1+\alpha}+\beta BW_{\rho'}}{1+\beta}
\ee
The mass and the width of the $\rho$ as well as $\alpha$ and $\beta$  
were free parameters of the fit, while the other paramaters were kept fixed 
(to the  values of~\cite{Akhmetshin:2001ig}).

The results obtained from the fit were: $M_{\rho} = (772.6\pm0.5)\,MeV$, 
$\Gamma_{\rho} = (143.7\pm0.7)\,MeV$, 
$\alpha = (1.48\pm0.12)\cdot10^{-3}$, $\beta = -0.147\pm0.002$.
These values, even if preliminary,  are in good agreement with 
the ones found by CMD-2~\cite{Akhmetshin:2001ig} 
 using a light different parametrization (Gounaris-Sakurai).  This agreement can be also confirmed  
from Fig.~\ref{fig:pion}, which shows 
the results of the two  parametrization (KLOE {\it blue line}, 
CMD-2 {\it red line}) overimposed to our experimental data. 
The discrepancy at low $Q^2_{\pi\pi}$ could be due either to the model 
inadequacy or to residual background events.

\section{Conclusion and outlook}
The status of $e^{+}e^{-} \rightarrow \ppg$ analysis and the preliminary results obtained on the pion form factor
were presented. 
The analysis was based on an integrated luminosity of $\sim73\,pb^{-1}$ , which corresponds to about 1/3 of the 
data collected (and already reconstructed) in 2001. 
The average error for single point above $0.4\, GeV^2$
is $\sim 2\%$, mainly dominated by the the limited Monte Carlo statistics in the efficiency evaluation.  
FSR and other background channels (as for example Bhabha's or $\pi^+\pi^-\pi^0$) are kept below 1\%  
by a likelihood method and kinematical cuts of the analysis. 
Other sources of systematical errors  due to luminosity, 
radiative corrections and resolution
 are  still under study, and should be included into our errors.

A comparison of the pion form factor with the fitted parametrization obtained by the CMD-2 collaboration,
shows a good (preliminary) agreement, even without unfolding the spectrum for the detector resolution effects.
Such a comparison, even at an accuracy of 2-3\%, has become more and more important
 in view of the $1.6\,\sigma$ discrepancy between 
$e^+e^-$ and $\tau$ data in the recent evaluation of $a^{had}_{\mu}$~\cite{Davier:2002dy}.
  
In order to improve the accuracy on  $a^{had}_{\mu}$ 
a final precision for this measurement below $1\%$ is needed. This is very important both in the energy region
 around the $\rho$, where the  CMD-2  collaboration has recently reached a 0.6\%  accuracy  as well as 
below $0.6\,GeV$, which contributes to $a_{\mu}$  with about $100\times 10^{-10}$ and is known with a worse  
accuracy~\cite{eidelman}.
KLOE can study both these regions using $\ppg$ events with the hard photon emitted at small and large angle. 
However in order to reach such an highly demanding  accuracy many factors must be well under control, 
both experimentally and  theoretically. The results here presented, together with the 
larger statistics up to now collected by KLOE  ($\sim500\, pb^{-1}$) and the intense theoretical
work on radiative corrections from different groups, are a promising indication for achieving such a challenging task.


\end{document}